\documentclass[journal]{IEEEtran}

\usepackage{amsmath,amssymb}
\usepackage{graphicx}
\usepackage{booktabs}
\usepackage{cite}
\usepackage{url}
\usepackage{xcolor}
\usepackage{tcolorbox}

\graphicspath{{figures/}}

\newcommand{\dnl}{d_{\mathrm{NL}}}

\begin{document}

\title{How Big Should a Wireless Foundation Model Be?}

\author{
\IEEEauthorblockN{Wei-Lun Cheng and Wanjiun Liao\\
\IEEEauthorblockA{Department of Electrical Engineering,
National Taiwan University, Taipei, Taiwan}\\
Email: \{d11921b15, wjliao\}@ntu.edu.tw}
}

\maketitle

{\noindent\small This work has been submitted to the IEEE for possible publication. Copyright may be transferred without notice, after which this version may no longer be accessible.\par}
\vspace{6pt}

\begin{abstract}
Wireless foundation models are rapidly emerging as a key enabler
of AI-native communication systems, yet a fundamental question
remains unanswered: how large should these models be?
We present a principled, physics-grounded answer, showing that the
\emph{intrinsic dimensionality} ($d_\mathrm{NL}$, the nonlinear
manifold dimension of the channel) acts as the
fundamental bottleneck, defining the scaling ceiling once a
data-sufficient regime is reached.
This dimensionality is not a design choice but a physical
constraint: Maxwell's equations, finite scatterers, and antenna
aperture inherently constrain wireless propagation environments
to a limited number of degrees of freedom---spanning 5--35
across both real-world OTA measurements and 3GPP-standardized
channel models we evaluate---orders of magnitude below
the ${\sim}$1,000-dimensional semantic space of language.
As a consequence, we propose a \emph{scaling framework} for
wireless AI: taking NTN satellite channels as a representative
case ($d_{\mathrm{NL}}\!\approx\!14$), scaling gains diminish
rapidly beyond ${\sim}$30 million parameters, entering a
stochastic asymptote above 70M where a further
1.6$\times$ increase (96M$\to$150M) yields only 0.52\,dB.
Beyond this ceiling, inference-time adaptation via pilot-aided
test-time training (TTT) is far more effective: a compact
12M-parameter model surpasses a static 96M model by
9.9\,dB (NMSE, SNR$\,=\,$20\,dB) / 7.6\,dB (MCM, SNR$\,=\,$10\,dB)
at one-eighth the parameters.
With $d_{\mathrm{NL}}$ distributions validated across
real-world indoor massive MIMO measurements, our scaling laws
and TTT gains are demonstrated through NTN satellite simulations,
reframing wireless AI design:
\emph{channel geometry---not model size---fundamentally governs
the scaling laws of physical-layer wireless AI}.
\end{abstract}

\begin{IEEEkeywords}
Wireless foundation model, scaling laws, test-time training,
intrinsic dimensionality, 6G.
\end{IEEEkeywords}

\section{Introduction}
\label{sec:intro}

Wireless foundation models are scaling rapidly.
Large Wireless Model (LWM)~\cite{alkhateeb2024lwm} deploys 85M
parameters; WiFo~\cite{wifo2025} applies masked autoencoders at
${\sim}$50M; WiMamba~\cite{wimamba2026} pursues architectural
efficiency; Tiny-WiFo~\cite{tinywifi2025} distills to 5.5M.
Each chooses its model size by reasoning about hardware
constraints, computational budgets, or analogy to NLP---yet a
more fundamental question \textbf{remains underexplored: what does
the channel actually require?}

The answer, we will show, is rooted in physics rather than
engineering.
Wireless channels are not arbitrary high-dimensional
distributions; they are constrained geometric objects, shaped by
a finite number of scatterers, the aperture of the antenna array,
and the laws of electromagnetic propagation.
When we measure channel complexity using non-linear intrinsic
dimensionality estimators---tools that see the curved manifold
structure that PCA misses---every channel environment we examine
falls in the range of 5--35 effective dimensions.
Language occupies roughly 1,000.
This gap is the central fact of wireless AI scaling, and this
paper identifies a governing principle.

\begin{figure}[t]
\centering
\includegraphics[width=\columnwidth]{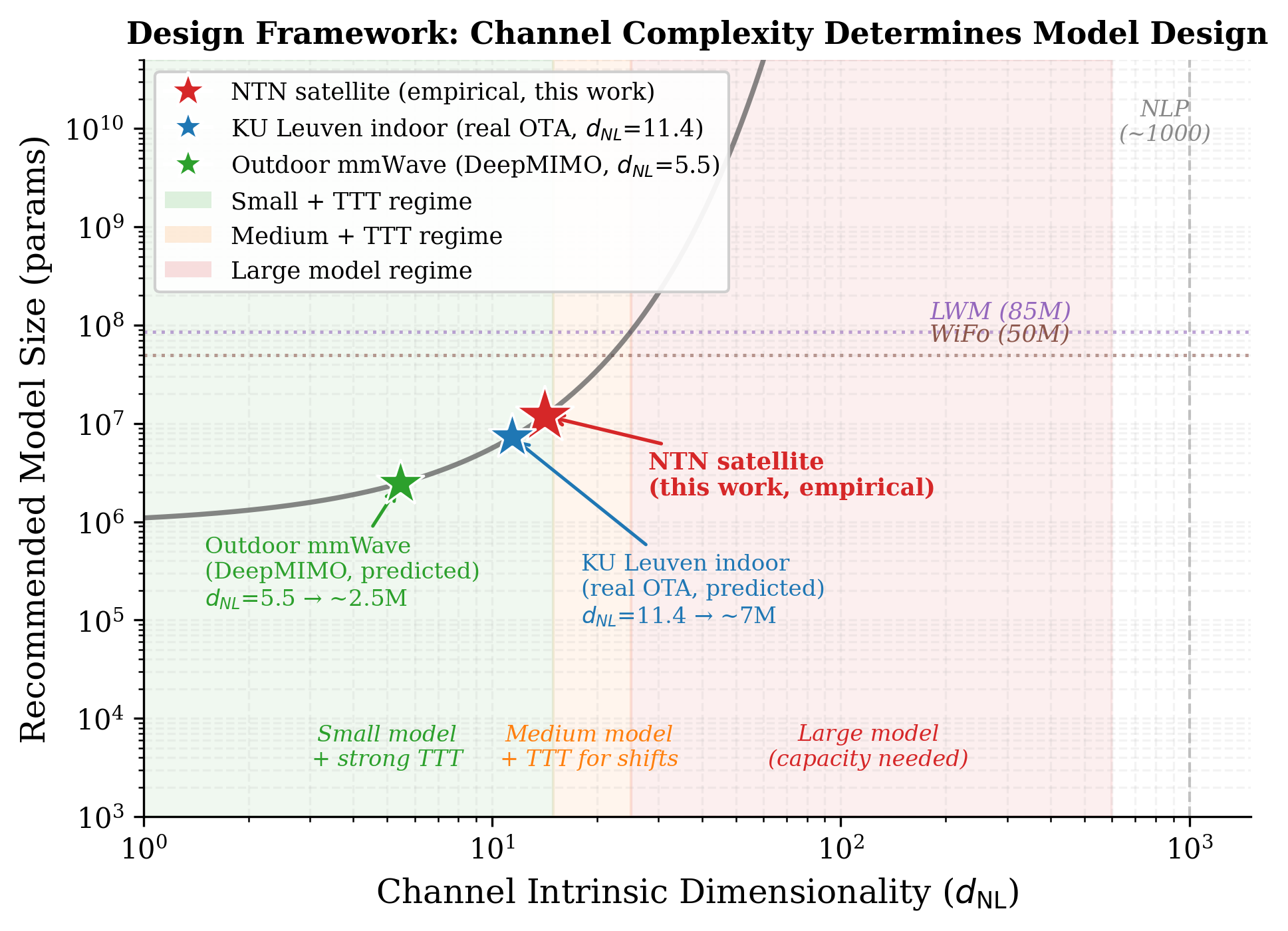}
\caption{Dimensionality-guided design framework.
Operating points for NTN satellite ($\dnl\approx14$,
this work, ${\sim}$12M, measured), KU~Leuven indoor OTA
($\dnl\approx11.4$, ${\sim}$7M, est.), and outdoor mmWave
($\dnl\approx5.5$, ${\sim}$2.5M, est.\ from scaling law)
all fall in the small-to-medium model regime.
Existing large models (LWM 85M, WiFo 50M) significantly
overshoot the complexity of any measured wireless channel.
Background zones indicate the recommended design strategy:
small model with aggressive TTT (green), medium model with
TTT for domain shifts (orange), large model for very high
complexity channels (red).}
\label{fig:framework}
\end{figure}

In this work, we answer this question through the first empirical
scaling law for wireless foundation models and propose a design
framework that connects \emph{channel intrinsic complexity} to
\emph{optimal model design} (Fig.~\ref{fig:framework}).
Our key insight is that the effective complexity of the propagation
environment determines (i)~how quickly model scaling saturates and
(ii)~how much value test-time adaptation provides.

Through validation across multiple channel domains---including
satellite, terrestrial indoor, outdoor mmWave, and five
standardized 3GPP CDL models---we show that wireless foundation
models exhibit significantly steeper scaling behavior than NLP
models, but also saturate at much smaller sizes.
This observation motivates a shift from ``scale-at-all-costs''
toward \emph{right-sized models with adaptive capabilities}.

To enable such adaptability, we introduce pilot-aided test-time
training (TTT), a lightweight mechanism that leverages standard
pilot signals to refine model predictions during deployment.
Rather than relying solely on large pre-trained models, TTT allows
smaller models to dynamically adjust to changing channel
conditions, bridging the gap between model capacity and
environmental variability.

\section{Why Wireless Is Fundamentally Different from NLP}
\label{sec:why}

In NLP, language has high intrinsic complexity---roughly a thousand
independent semantic dimensions~\cite{aghajanyan2021intrinsic}---which
is why trillion-parameter models continue to improve.
Wireless channels are governed by different physics.
The propagation environment is constrained by a finite number of
scatterers, antenna aperture, and bandwidth, yielding far fewer
independent degrees of freedom.

\subsection{Measuring Channel Complexity}

How complex is a wireless channel?
The traditional answer uses PCA (principal component analysis):
count how many linear components capture 90--95\% of the variance.
For satellite channels, PCA says 42; for outdoor mmWave, 191.
These numbers suggest high complexity.

But PCA only sees \emph{linear} structure.
Wireless channels have inherently non-linear structure---steering
vectors are complex exponentials, phase varies on a circle, and
delay-angle coupling creates curved manifolds.
When we apply non-linear estimators (Two-NN~\cite{facco2017twonn},
validated against MLE~\cite{levina2005mle}), the picture changes
dramatically.
We denote the resulting estimate $\dnl$ (non-linear intrinsic
dimension).
All $\dnl$ values below are estimated from channel matrices
observed at a $N_\mathrm{BS}\!=\!128$-antenna BS with 76
OFDM subcarriers---the configuration used throughout this work.
Varying the array size shifts absolute $\dnl$ values but
preserves the relative ordering across environments.

\begin{itemize}
\item \textbf{Satellite (NTN):} PCA says 42 dimensions;
$\dnl \approx 14$---varying from 8 at zenith (line-of-sight
dominant) to 22 at low elevation (rich multipath)~\cite{3gpp38811}.
\item \textbf{Indoor terrestrial:} Real over-the-air measurements
from the KU Leuven massive MIMO sounder~\cite{kuleuven2024mimo}
confirm $\dnl \approx 10$--$15$, consistent with our
simulation-based predictions.
\item \textbf{Outdoor mmWave:} PCA says 191; non-linear analysis
says only ${\sim}$5.5.
The 35$\times$ gap arises because mmWave steering vectors are
non-linear functions of angle---what appears as 191 linear
dimensions is actually a 5.5-dimensional curved manifold.
\item \textbf{3GPP CDL models:} Across five standardized channel
models (CDL-A through CDL-E), $\dnl$ ranges from 12 (strong LoS)
to 35 (rich NLOS)---perfectly tracking the physical scattering
complexity~\cite{3gpp38901}.
\end{itemize}

\begin{figure}[t]
\centering
\includegraphics[width=0.9\columnwidth]{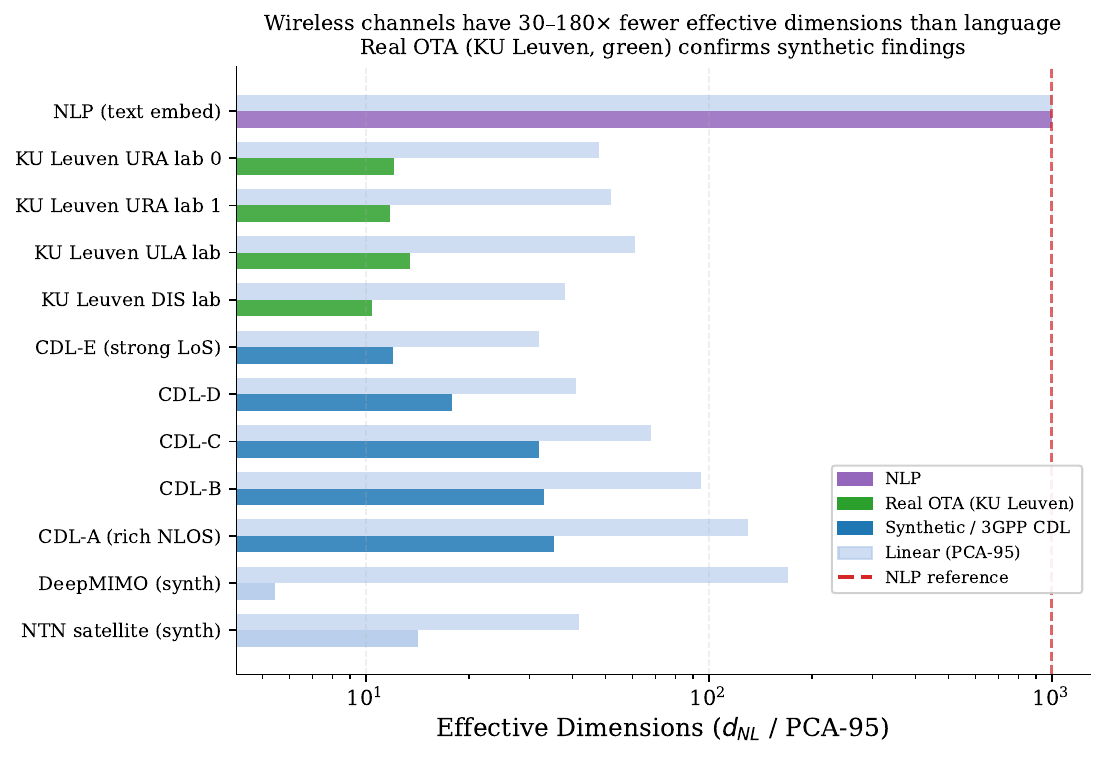}
\caption{Intrinsic complexity across data domains.
Wireless channels have 5--100$\times$ fewer effective degrees of
freedom than language, explaining why wireless models saturate at
much smaller sizes.
The non-linear estimate (Two-NN) reveals that even the PCA values
significantly overstate the true channel complexity.}
\label{fig:deff}
\end{figure}

These measurements are summarized in Fig.~\ref{fig:deff}.
The message is clear: \emph{all} wireless channels have intrinsic
complexity far below language.
This is the fundamental reason why wireless foundation models
should not follow the NLP scaling playbook.

\subsection{What This Means for Scaling}

Low intrinsic complexity has a direct consequence for model
scaling: models learn the channel structure quickly, and further
scaling yields diminishing returns.

We trained masked autoencoder models at twelve scales spanning
three orders of magnitude (104K to 150M parameters) on 224,000
NTN satellite channel matrices.
Three scaling phases emerge along the parameter axis:

\begin{enumerate}
\item \textbf{Rapid learning} (104K--12M): Each 10$\times$
parameter increase yields ${\sim}$4\,dB improvement in
reconstruction quality.
The model-size scaling exponent in this regime ($\alpha \approx 0.41$)
is roughly five times steeper than NLP's
$\alpha \approx 0.076$~\cite{kaplan2020scaling}.
\item \textbf{Diminishing returns} (12M--96M): Gains shrink
overall, though not monotonically---discrete jumps in model
architecture (layer depth, embedding dimension) create some
variation between adjacent scales.
From 12M to 96M (an 8$\times$ increase), the cumulative gain
in MCM reconstruction loss is only ${\sim}$2.3\,dB
(and merely ${\sim}$0.8\,dB in downstream NMSE at
SNR$\,=\,$20\,dB, Table~\ref{tab:comparison}).
\item \textbf{Saturation} ($>$96M): A 1.6$\times$ parameter
increase from 96M to 150M yields only 0.52\,dB---effectively
zero improvement, confirming that the channel manifold's
learnable structure has been exhausted.
\end{enumerate}

This pattern is consistent across domains.
On outdoor mmWave data (DeepMIMO~\cite{deepmimo2022}), scaling
saturates near 12M parameters---consistent with its low $\dnl$
of only ${\sim}$5.5---while on real measured terrestrial channels
(DICHASUS~\cite{dichasus2022}, 578 samples), saturation appears
even earlier, though the small dataset size makes it difficult to
disentangle data limitation from physical saturation in this case.

The practical implication is stark: scaling from 12M to 96M
costs 8$\times$ more silicon and power, yet delivers only
2.3\,dB in MCM quality (0.8\,dB in NMSE)---less than one-ninth
of the 7.2\,dB that 5~TTT steps provide in NMSE.

The effective scaling window closes near ${\sim}$100M
parameters---a stark contrast to NLP, where the window extends
to hundreds of billions.

\begin{figure}[t]
\centering
\includegraphics[width=\columnwidth]{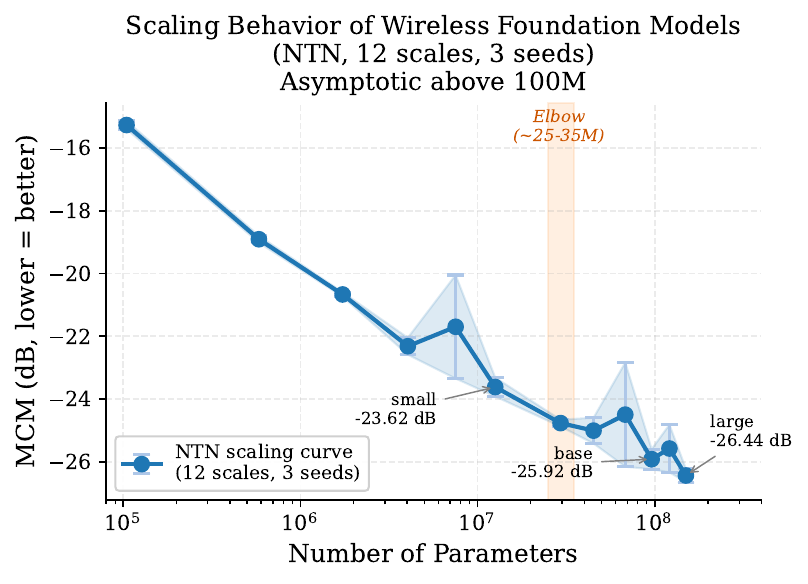}
\caption{Scaling behavior of wireless foundation models on
NTN satellite channels (12 scales, unified BS$=$128).
Performance improves rapidly at small model sizes but enters
diminishing returns beyond ${\sim}$10M parameters, with
effective saturation at ${\sim}$100M (base$\to$large $=$
$+$0.52\,dB).
This early saturation reflects the limited intrinsic complexity
of wireless channels---in stark contrast to NLP, where scaling
continues to yield gains at billions of parameters.}
\label{fig:scaling}
\end{figure}

\section{Test-Time Training: Adapt, Don't Scale}
\label{sec:ttt}

Since scaling beyond ${\sim}$10M parameters yields diminishing
returns, we ask: \emph{what is a better use of compute?}

Our answer is \textbf{test-time training (TTT)}: adapting the
model at inference time to the current channel conditions.
Conceptually, TTT is analogous to re-calibrating an adaptive
equalizer using known training sequences---but operating on the
neural network weights rather than filter taps.

\subsection{Leveraging Free Supervision}

In standard OFDM systems, pilot symbols are inserted at known
subcarrier positions in every transmitted frame.
These pilots are ``free'' self-supervised labels: the receiver
knows what was sent and can compare it against what the model
predicts.

TTT exploits this by performing a few gradient descent steps on
the decoder parameters, minimizing the reconstruction error at
pilot positions.
Only the lightweight decoder (${\sim}$11\% of the model) is
updated; the encoder---which stores the learned channel
prior---is frozen.
Parameters are reset when the channel scenario changes.

\subsection{Small + Adaptive Beats Large + Static}

Table~\ref{tab:comparison} summarizes the key comparison.
A 12M model with 5 TTT steps outperforms its own static inference
by 7.2\,dB in channel estimation NMSE at SNR$\,=\,$20\,dB---using
only standard LS-estimated pilots with 12.5\% overhead (every 8th
subcarrier).

\begin{table}[t]
\centering
\caption{Why adapt instead of scale. Gains are reported at
SNR$\,=\,$20\,dB (NMSE) and noiseless (MCM).
Adapt.\ GFLOPs: one-time TTT cost (amortised across $N\!>\!2$
subsequent inferences); Infer.\ GFLOPs: per-inference cost in deployment.
At SNR$\,=\,$10\,dB the same 12M+TTT model provides
4.75\,dB (5~steps) / 7.23\,dB (20~steps) NMSE self-improvement;
the abstract reports 7.6\,dB MCM gain vs.\ static 96M at
SNR$\,=\,$10\,dB.}
\label{tab:comparison}
\begin{tabular}{lrcc}
\toprule
\textbf{Strategy} & \textbf{Gain} & \textbf{Adapt.G} & \textbf{Infer.G} \\
\midrule
Scale 12M$\!\to\!$96M & ${\sim}$2.3\,dB (MCM)        & ---  & 0.9 \\
Scale 12M$\!\to\!$96M & ${\sim}$0.8\,dB (NMSE@20\,dB) & ---  & 0.9 \\
Adapt 12M+5\,TTT      & \textbf{7.2\,dB} (NMSE@20\,dB)  & 1.7  & 0.1 \\
Adapt 12M+20\,TTT     & \textbf{10.8\,dB} (NMSE@20\,dB) & 6.8  & 0.1 \\
\bottomrule
\end{tabular}
\end{table}

Under the same NMSE metric, scaling from 12M to 96M yields
only ${\sim}$0.8\,dB---while 5~TTT steps deliver 7.2\,dB,
a $9{\times}$ performance gap.
The compute picture is equally striking: the 96M model costs
0.9\,GFLOPs \emph{every} inference, whereas the adapted 12M
model costs just 0.1\,GFLOPs per inference---a
$9{\times}$ reduction in sustained deployment cost.
The one-time TTT adaptation (1.7\,GFLOPs for 5~steps)
breaks even after only ${\sim}$2 inferences, after which
the smaller model is both cheaper and more accurate.

Fig.~\ref{fig:perf}(a) shows the convergence behavior: performance
improves monotonically with TTT steps and saturates at
${\sim}$20 steps with no sign of overfitting---the model
gracefully converges rather than memorizing noisy pilots.
This robustness arises because only the lightweight decoder
(${\sim}$11\% of parameters) is updated while the encoder is
frozen, limiting capacity available to overfit; LS pilots,
though noisy, carry consistent geometric structure across the
adaptation window, so gradient descent converges to channel
geometry rather than noise realizations.
Even 5--10 steps capture 65--80\% of the full gain.

Pre-training is essential: a randomly initialized model shows
\emph{zero} TTT improvement (Fig.~\ref{fig:perf}(a), dashed).
The 14\,dB gap between pre-trained and random confirms that
pre-training encodes a structural prior of the channel manifold
that TTT alone cannot discover.

\subsection{Cross-Domain Transfer}

Can a model trained on satellite data adapt to terrestrial
channels?
We tested our NTN-trained model on real measured indoor channels
from DICHASUS~\cite{dichasus2022}---an extreme domain shift
(Ka-band satellite to sub-6\,GHz indoor, different antenna
geometry, different propagation physics).
Both datasets are preprocessed to a common tensor shape via
subcarrier selection and antenna-port resampling; the MAE
tokenizes channel matrices into fixed-size patches, so no
architectural modification is needed.

Without TTT, the model fails completely ($-0.15$\,dB NMSE, near
the noise floor).
With 20 TTT steps, it recovers to $-6.49$\,dB---a
\textbf{6.34\,dB gain} that restores a viable link from the
noise floor using only pilot observations, with no
target-domain labels.

\begin{figure}[t]
\centering
\includegraphics[width=\columnwidth]{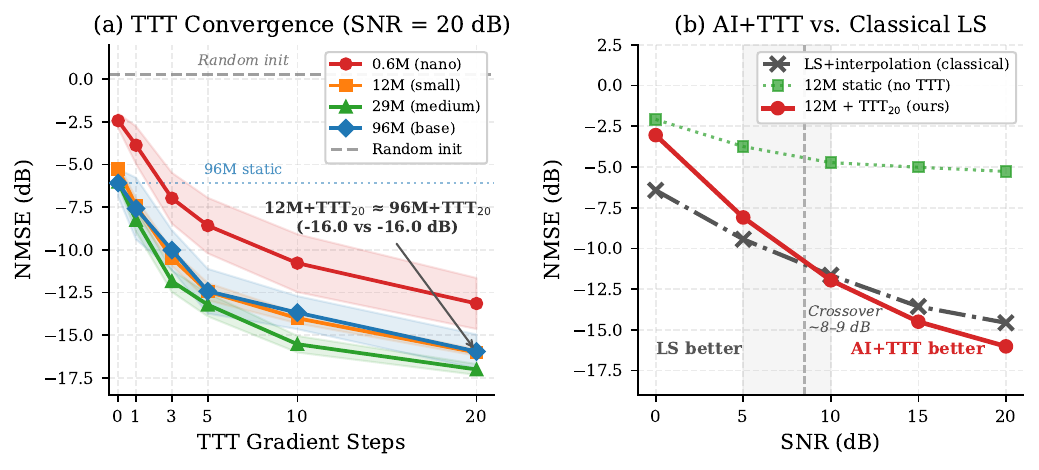}
\caption{Test-time training performance.
\textbf{(a)}~Convergence across model scales (SNR$\,=\,$20\,dB):
performance improves monotonically and saturates at ${\sim}$20 steps
with no overfitting; a randomly initialized model shows zero TTT
improvement, confirming pre-training is essential.
Even 5--10 steps capture 65--80\% of the full gain.
\textbf{(b)}~NMSE vs.\ SNR: 12M+TTT$_{20}$ vs.\ classical
LS+interpolation (same 12.5\% pilot overhead).
A crossover near SNR$\,\approx$\,8--9\,dB divides two regimes:
LS is more robust at low SNR; AI+TTT outperforms at high SNR
where interpolation error is the binding constraint.}
\label{fig:perf}
\end{figure}

For context, $-6.49$\,dB NMSE corresponds to an estimation error
of ${\sim}$22\% of the channel energy---sufficient for robust
low-rate signaling (e.g., QPSK) but not for high-order modulation
(e.g., 64-QAM).
The key observation is that this rapid recovery is only possible
because pre-training encodes a structural prior of the channel
manifold---the same conclusion demonstrated in the in-domain
setting (Fig.~\ref{fig:perf}(a), dashed): a randomly initialized
model cannot discover channel structure from pilots alone.
Full domain-specific training remains preferable when data is
available; the value of cross-domain TTT lies in rapid emergency
adaptation when it is not.

\subsection{The Practical Operating Regime}

Fig.~\ref{fig:perf}(b) compares our 12M+TTT model against classical
LS+interpolation (every 8th subcarrier, same 12.5\% pilot overhead).
The results reveal a clear crossover near SNR\,$\approx$\,8--9\,dB.

At \textbf{low SNR} ($<$8\,dB), LS+interpolation is more robust,
outperforming AI+TTT by up to 3.4\,dB at 0\,dB SNR.
The reason is intuitive: under heavy noise, the TTT process adapts
to noise artifacts rather than the true channel, while LS averaging
over multiple pilots suppresses noise through linear smoothing.

At \textbf{high SNR} ($>$10\,dB), AI+TTT surpasses classical
methods by up to 1.5\,dB at SNR\,$=\,$20\,dB.
With a clean signal, the adapted model captures non-linear
channel structure that linear interpolation cannot resolve.

This crossover defines a \textbf{hybrid deployment rule}:
use classical LS+interpolation for SNR$\,<\,$8\,dB;
activate AI+TTT for SNR$\,>\,$10\,dB.
In practice, AI+TTT is most valuable in high-SNR cell-center
and NTN LoS scenarios---where its manifold-aware reconstruction
outperforms the interpolation ceiling of classical methods.

\section{A Design Framework for Wireless AI}
\label{sec:framework_detail}

Our findings lead to a practical design framework:
\emph{the intrinsic complexity of the target channel determines
the optimal balance between model capacity and inference-time
adaptation.}

\subsection{How Existing Approaches Compare}

Table~\ref{tab:landscape} positions our approach relative to
existing wireless foundation models.
The key distinction is between approaches that pursue scale
(LWM, WiFo) and our approach that pursues \emph{right-sizing
guided by channel physics}.
Tiny-WiFo shares our goal of smaller models but achieves it
through knowledge distillation from a large teacher---which
still requires training the large model first.
Our approach skips the large model entirely: measure $\dnl$,
size accordingly, and adapt with TTT.

\begin{table}[t]
\centering
\caption{Wireless foundation model landscape: different strategies
for the same question.}
\label{tab:landscape}
\begin{tabular}{lcccc}
\toprule
\textbf{Approach} & \textbf{Params} & \textbf{Adapt.}
& \textbf{Strategy} & \textbf{Sizing basis} \\
\midrule
LWM~\cite{alkhateeb2024lwm} & 85M & Zero-shot & Scale up
& Hardware limit \\
WiFo~\cite{wifo2025} & ${\sim}$50M & None & Scale up
& Intuition \\
Tiny-WiFo~\cite{tinywifi2025} & 5.5M & None & Distill
& Teacher size \\
WiMamba~\cite{wimamba2026} & Various & None & Eff.\ arch.
& Complexity \\
\textbf{Ours} & \textbf{12M} & \textbf{TTT}
& \textbf{Right-size} & \textbf{Channel $\dnl$} \\
\bottomrule
\end{tabular}
\end{table}

The key distinction: existing approaches choose model size based
on hardware constraints, intuition, or the teacher model's
capacity.
Our approach is the first to derive model size from the
\emph{physical complexity of the channel itself}.
Tiny-WiFo shares our goal of smaller models but requires training
a large teacher first---our approach skips the large model
entirely.

\subsection{A Practitioner's Recipe}

For engineers deploying wireless foundation models:

\begin{enumerate}
\item \textbf{Profile the channel.}
Collect ${\sim}$5,000 representative channel samples and measure
$\dnl$ using the Two-NN estimator.
This requires no GPU and takes minutes on a laptop.
A useful empirical diagnostic: compute $\rho = N / 2^{\dnl}$
where $N$ is the training set size.
Across all datasets we examine, $\rho \ll 10^{-4}$ predicts
degenerate collapse (all model sizes fail) and $\rho > 10$
enables healthy power-law scaling; the intermediate regime
exhibits degraded or saturated performance.
This quantity serves as a practical collapse-risk indicator,
grounded in the exponential covering cost of high-dimensional
manifolds; its full theoretical treatment is beyond the scope
of this article, but the empirical correlation is consistent
across all ten channel datasets we examine (NTN, KU~Leuven, RENEW~OTA, DICHASUS, CDL-A/B/C/D/E, DeepMIMO).
\item \textbf{Size the model.}
$\dnl < 15$ (satellite, strong LoS): 10--15M parameters.
$\dnl \approx 15$--$25$ (terrestrial indoor): 15--50M.
$\dnl > 25$ (rich NLOS urban): 30--100M \emph{only when
data is abundant} ($\rho \gg 1$); when per-domain data is
scarce, large models may suffer degenerate collapse---revert
to small models (${\leq}$15M) with multi-domain pre-training
or TTT.
In all cases, size to channel physics, not hardware budget.
\item \textbf{Deploy with TTT.}
Allocate 5--10 pilot-aided gradient steps per adaptation episode.
This captures 65--80\% of the full TTT gain at a fraction of the
compute of 20 steps.
\item \textbf{Know when NOT to use a foundation model.}
For simple channels ($\dnl < 10$, strong LoS with high Rician
$K$-factor), classical LMMSE with a single covariance estimate
is simpler and equally effective.
\end{enumerate}

\subsection{Worked Example}

Consider an NTN operator deploying at Ka-band with satellite
elevation ranging from $30^\circ$ to $90^\circ$.
Step~1: Collecting 5,000 channel samples and running Two-NN
(3~minutes on a laptop) yields $\dnl \approx 14$.
Step~2: The sizing guideline suggests 10--15M parameters.
Step~3: A 12M model is trained and deployed with 5--10~TTT steps per
elevation change.
Expected performance: ${\sim}$$-10$\,dB NMSE at SNR$\,=\,$10\,dB
(5~steps: $-9.5$\,dB; 10~steps: $-10.5$\,dB)---comparable to a
static 96M model at $9{\times}$ lower inference cost.
Total deployment: 25\,MB model weights (FP16), ${\sim}$100\,MB
working memory, $<$0.1\,s adaptation latency per scenario change.

\subsection{Beyond Channel Estimation}

While we validate our framework on channel estimation, the
underlying principle extends further.
The intrinsic complexity $\dnl$ characterizes the channel
\emph{data manifold}---not the estimation task.
Any model operating on channel data---for beamforming, CSI
feedback, or positioning---is constrained by the same manifold
complexity.

This argument has a rigorous foundation: channel estimation
reconstructs the \emph{entire} manifold, making it the most
information-demanding task on channel data.
By the data processing inequality, any downstream task that
extracts lower-dimensional features---such as beam index
prediction (output: one integer) or positioning (output: 3D
coordinates)---requires strictly less model capacity.
If models saturate at ${\sim}$10--30M for full manifold
reconstruction, downstream tasks can only saturate \emph{earlier}.

We therefore expect the design guidelines above to apply broadly,
though the precise scaling exponent may vary across tasks.
Validating this cross-task prediction is an important open
direction.

\section{Open Challenges and Future Directions}
\label{sec:challenges}

\subsection{Algorithmic Frontiers}

\textbf{Cross-architecture validation.}
Since $\dnl$ is a data property, the saturation should hold for
any architecture.
Preliminary CNN experiments show consistent behavior, but broader
validation across Mamba and hybrid architectures is needed.

\textbf{Cross-task scaling.}
Does the exponent differ for beamforming, positioning, or CSI
feedback?
Channel estimation directly reconstructs the manifold, which may
represent an upper bound on scaling efficiency; tasks with
lower-dimensional outputs could saturate even earlier.

\textbf{Compute-optimal training.}
Jointly optimizing model size and dataset size for a fixed compute
budget---following the Chinchilla methodology for
wireless~\cite{hoffmann2022chinchilla}---would yield more precise
deployment guidance, completing the picture from ``how big'' to
``how big, trained on how much.''

\subsection{Why Wireless Cannot Follow the NLP Playbook}

A deeper question is \emph{why} NLP can keep scaling while
wireless cannot.
Language complexity arises from the unbounded combinatorics of
human semantics---there is no physical ceiling on how many
independent meanings a sentence can carry.
Wireless channel complexity, by contrast, is locked by Maxwell's
equations and spatial geometry: a finite number of scatterers,
a fixed antenna aperture, and a bounded bandwidth yield a
physically immutable upper limit on independent degrees of freedom.
\emph{NLP learns infinite culture; wireless AI learns finite
geometry.}

\subsection{Hardware and Deployment Challenges}

\textbf{Hardware-software co-design.}
TTT requires backpropagation through the decoder, but current
edge NPUs and baseband ASICs are designed exclusively for forward
inference---they lack the gradient storage and dynamic memory
allocation needed for weight updates.
Enabling on-device TTT will require future 6G chipsets to support
lightweight, decoder-only gradient computation.
Until then, TTT is best suited for GPU-equipped base stations
rather than user devices.

\textbf{Real-time adaptation.}
Our TTT operates at the scenario-adaptation timescale (seconds to
minutes), not the fast-fading timescale (${\sim}42\,\mu$s for NTN).
For NTN at 20\,GHz, the profiled cost of 5~TTT steps is
1.7\,GFLOPs, corresponding to ${\approx}34$\,ms on an ARM-class
CPU (50\,GFLOPS)---well within the LEO coherence interval
(${\ge}100$\,ms) and negligible relative to the
seconds-level scattering geometry timescale.
For vehicular channels (coherence time ${\sim}1$\,ms),
single-step TTT at ${\approx}7$\,ms is too slow for fast-fading
tracking; TTT is best suited to geometry-level changes
(e.g., scenario transitions) that evolve on second-level
timescales.

\textbf{Multi-user scalability.}
A base station serving hundreds of users in different scattering
environments must maintain per-user decoder states.
Clustering users with similar channel geometry to share TTT
weights, or applying low-rank adaptation (LoRA), could
significantly reduce the per-user memory overhead.

\subsection{Ecosystem Implications for 6G}

\textbf{Data collection strategy.}
Since $\dnl$ is only 5--35, models quickly exhaust the variance
in routine channel conditions.
Future data collection should shift from brute-force accumulation
of petabytes of common scenarios toward \emph{active learning}
that targets rare edge cases---a fundamentally different ROI
calculus for operators investing in channel measurement campaigns.

\textbf{Model orchestration.}
Dimensionality-guided sizing enables a new deployment paradigm:
rather than pushing a single massive model to every network node,
a central management plane can profile each cell's channel
complexity (via $\dnl$) and dynamically dispatch right-sized models
to each node, reducing backhaul bandwidth and edge power consumption.
This principle is agnostic to any specific RAN architecture.
In open, disaggregated deployments, this logic maps naturally onto
the RAN Intelligent Controller; in proprietary networks, it can be
implemented in any centralized network management function.

\textbf{Implications for standardization.}
Current 3GPP pilot patterns were designed for linear estimators
such as LMMSE.
If AI-native channel estimation with TTT becomes the 6G paradigm,
future standardization should explore pilot distributions optimized
for gradient-based adaptation---rethinking physical-layer reference
signals as training inputs rather than mere interpolation anchors.

\section{Conclusion}
\label{sec:conclusion}

The wireless AI community has inherited an assumption from NLP:
that larger models are always better.
This article challenges that assumption.

The complexity of wireless channels is inherently limited by
physical propagation, resulting in much lower intrinsic
dimensionality than language or vision data.
As a consequence, wireless foundation models exhibit rapid gains
at small scales, followed by early saturation beyond which further
scaling is inefficient.

The right strategy is not to build bigger models, but to build
\emph{appropriately sized models that adapt at inference time}.
Pilot-aided test-time training makes this concrete: a few gradient
steps on standard OFDM pilots yield improvements equivalent to
significantly more hardware, at a fraction of the computational
cost.

For the channels that matter most---non-terrestrial, cell-edge,
rapidly varying---the path forward is not larger models, but
\emph{smarter ones}: right-sized to match channel geometry,
and equipped to adapt at inference time.
In the journey toward 6G AI-native networks, we believe that
measuring $\dnl$ before choosing model size will become as routine
as measuring coherence time before choosing modulation order---a
fundamental design parameter rooted in physics, not in hype.



\bibliographystyle{IEEEtran}

\end{document}